\documentclass[]{article}
\usepackage{proceed2e}
\usepackage{amsmath}
\usepackage{newalg}
\usepackage{amsfonts}
\usepackage{subfigure}
\usepackage{graphicx}
\usepackage{booktabs}

\providecommand{\tabularnewline}{\\}

\title{Communication Communities in MOOCs}

\author{ {\bf Nabeel Gillani, Michael Osborne, Stephen Roberts} \\
Machine Learning Research Group\\
Department of Engineering Science\\
University of Oxford\\
\{nabeel, mosb, sjrob\}@robots.ox.ac.uk \\
\And
{\bf Rebecca Eynon, Isis Hjorth}   \\
Oxford Internet Institute \\
Department of Education\\
University of Oxford \\
\{rebecca.eynon, isis.hjorth\}@oii.ox.ac.uk \\
}

\begin{document}

\maketitle

\begin{abstract}

Massive Open Online Courses (MOOCs) bring together thousands of people from different geographies and demographic backgrounds -- but to date, little is known about how they learn or communicate. We introduce a new content-analysed MOOC dataset and use Bayesian Non-negative Matrix Factorization (BNMF) to extract communities of learners based on the nature of their online forum posts. We see that BNMF yields a superior probabilistic generative model for online discussions when compared to other models, and that the communities it learns are differentiated by their composite students' demographic and course performance indicators. These findings suggest that computationally efficient probabilistic generative modelling of MOOCs can reveal important insights for educational researchers and practitioners and help to develop more intelligent and responsive online learning environments.
\end{abstract}

\section{Introduction}

There has been no shortage of polarised media hype around Massively Open Online Courses (MOOCs) -- and yet, grounded research on their pedagogical effectiveness and potential is in its infancy. Research in online education is not new: proponents have lauded its potential and critics have warned of an imminent ``mechanization of education'' since the popularization of the internet nearly two decades ago (Nobel, 1998).  Moreover, much education literature has focused on the way learners use online discussion forums, drawing upon social constructivist (Vygotsky, 1978) conceptions of knowledge creation and sharing.

The rise of MOOCs has provided a new opportunity to explore, among other things, how communication unfolds in a global, semi-synchronous classroom.  Still, early research efforts have taken only a cursory look at the nature of this communication, for example, detecting and modelling the prevalence of chatter irrelevant to the course (Brinton et al., 2013) and, in some cases, exploring the frequency of words used by those that pass or fail (Anderson et al., 2014).  Other studies in online education more broadly have leveraged social network analysis of fine-grained digital trace data to understand the nature of information cascades between groups of students (Vaquero and Cebrian, 2013).

In order for educators to support dialogue and interaction that enables learning in online settings, a theoretically and practically sound effort is needed to understand the different characteristics -- and emergent communities -- of learners that communicate with one another.  Fortunately, the problem of inferring communities, or more generally, latent features in datasets has been studied across domains (e.g., Holmes et al., 2012).  Indeed, some researchers have begun to explore how latent feature models like the mixed-membership stochastic block model (Airoldi et al., 2008) can be used to explain forum participation and dropout rates (Rose, 2013).  In these first attempts to make sense of the nature of interaction and learning in MOOC forums, the sheer volume of thousands of discussion forum posts has prevented latent feature analysis from modelling the \textit{content} and \textit{context} of posts -- attributes that must be considered in order to truly understand how people communicate and interact in global-scale learning settings.  

Our efforts aim to contribute to the expanding body of literature on MOOCs and the Machine Learning community by 1) introducing a new content-analysed dataset of MOOC forum data; 2) leveraging and validating Bayesian Non-negative Matrix Factorization as a robust probablistic generative model for online discussions; 3) illustrating how the tools of Machine Learning can not only reveal important insights into the nature of discussion in MOOCs, but how these tools may be leveraged to design more suitable and effective online learning environments.  

\section{Dataset and Latent Feature Models}

\subsection{Content Analysis of Forum Data}

Our primary research objective was to leverage Machine Learning to infer different groups of forum participants based on the content of their discussions, and to explore these groups with respect to user-specific digital trace data (such as their geographies and course outcomes) captured in the online course environment. 

We analysed data from a business strategy MOOC offered on the Coursera platform in Spring 2013. Nearly 90,000 students registered for the course, which lasted for six weeks and assessed students through a combination of weekly quizzes and a final project.  The online discussion forum was comprised of a number of sub-forums, which in turn had their own sub-forums or user-generated discussion threads that contained posts or comments.  There were over 15,600 posts or comments in the discussion forum, generated by nearly 4,500 learners.  Over 15,000 learners viewed at least one discussion thread in both instances, contributing to 181,911 total discussion thread views. 

We conducted qualitative content analysis on nearly 6,500 posts from this course -- to our knowledge, an unprecedented undertaking to date in MOOC research.  Content analyses have sometimes been used in online learning research, yet at much smaller scales than presented here (e.g.  De Weaver et al. 2006). The content analysis scheme for the present study was developed based on both existing academic literature and preliminary observations of online course discussions. 

We selected five dimensions to capture key aspects of interaction and learning processes.  The first dimension (learning) was used to collect data about the extent to which knowledge construction occurred through discussions, categorising each post using one of nine categories, ranging from no learning, through to four types of sharing and comparing of information, to more advanced stages of knowledge construction such as negotiation of meaning (Gunawardena et al., 1997).  The second dimension identified communicative intent in the forums, selecting from five categories:  argumentative, responsive, informative, elicitative and imperative (Clark et al., 2007; Erkens et al., 2008). The third dimension -- affect -- gauged levels and relative distributions of emotion in discourse, using five codes: positive / negative activating, positive / negative deactivating, and neutral (Pekrun et al., 2002).  Based on our own observations of the forums, we also developed two more dimensions: one related to topic, which had 11 categories that reflected all course related topics (e.g. cases, quizzes, readings, arrange offline meet-ups, introductions); and the other a rating of relevance of the post to its containing thread and sub-forum.  Relevance was rated on a three point scale: high relevance, low relevance and no relevance.    

For simplicity (given the size of the dataset), the unit of analysis selected was the post. The qualitative analysis software NVivo was used for labelling content.  Coding was conducted by four individuals who trained together over the course of two sessions and pilot tested the instrument together to enhance reliability. 

\subsection{Inferring Latent Features}
We are interested in probabilistic generative models to infer hidden features in content-analysed MOOC forum data for three reasons:  1) in a principled Bayesian setting, they enable the use of tools -- e.g. priors and likelihoods -- that are intuitively appropriate and fitting for the specific data at hand; 2) they do not require tuning resolution parameters or other values that govern the existence of communities (unlike many modularity algorithms -- e.g., Newman and Girvan, 2003); 3) they enable the simulation of clustered data, which, in the context of online education, may be particularly relevant to inform pedagogical practices, course design, and implementation.

One of the most widely-used latent feature models is Latent Dirichlet Allocation (LDA -- Blei et al., 2003), a generative model for inferring the hidden topics in a corpus of documents (e.g., articles in a newspaper).  A key attribute of LDA is that it represents a document's soft \textit{topic distribution}.  Blei et al. proposed variational inference for LDA, although subsequent extensions have proposed both collapsed and uncollapsed Gibbs samplers (e.g., Xiao and Stibor, 2010).  The Mixed Membership Stochastic Block (MMB -- Airoldi et al., 2008) is another generative process that models the soft-membership of items within a set of communities (or ``blocks''), further representing how different communities interact with one another.  The original MMB paper proposed a variational inference scheme based off of the one used for LDA, citing Markov Chain Monte Carlo (MCMC) methods as impractical given the large number of variables that would need to be sampled.    

In the Bayesian nonparametrics literature, the Indian Buffet Process (IBP -- Griffiths and Gharamani, 2005) is perhaps the most prominent infinite latent feature model, specifying a prior distribution over binary matrices that denote the latent features characterizing a particular dataset.  Similar to LDA and MMB, the IBP enables an observation to be characterized by multiple features -- but it treats the number of latent features as an additional variable to be learned as a part of inference.  Unfortunately, given the large (exponentially-sized) support over the distribution of possible binary matrices for any fixed number of features, the IBP has proven intractable for large-scale datasets.  The development of variational inference schemes and efficient sampling procedures (Doshi-Velez, 2009) over the past few years have enabled inference for simple conjugate (often linear-Gaussian) models with thousands of data points.  Recent explorations have revealed how inference in the IBP can be scaled via submodular optimization (Reed and Ghahramani, 2013), although again, by exploiting the structure of the linear-Gaussian likelihood model.  

While some have proposed models and inference schemes for the IBP with a Poisson likelihood (Gupta et al., 2012; Titsias, 2007), others have turned away from nonparametric models in favour of greater simplicity and computational efficiency.  Non-negative Matrix Factorization (Lee and Seung, 1999) is one such method used to produce a parts-based representation of an $N$ $\times$ $D$ data matrix $\mbox{X}$, i.e. $\mbox{X}$ $\approx$ $\mbox{W}$$\mbox{H}$, where $\mbox{W}$ and $\mbox{H}$ are $N$ $\times$ $K$ and $K$ $\times$ $D$ matrices, respectively.  A Bayesian extension to NMF (BNMF) was proposed by (Schmidt et al., 2008), which placed exponential priors on a Gaussian data likelihood model, presenting a Gibbs sampler (Geman and Geman, 1984) for inference.  While some of BNMF's applications have adopted fully Bayesian inference procedures, others have have pursued maximum likelihood or maximum a-posteriori (MAP) estimates, often for computational efficiency.

\vspace{-2mm}

\section{Bayesian Non-negative Matrix Factorization}

\vspace{-2mm}

In this work, we apply BNMF for community detection, as proposed by (Psorakis et al. 2011a) for social networks, to understand how latent communities can be inferred among MOOC users based on the content of their forum posts.  The model informs a \textit{belief} about each individual's community membership by presenting a membership distribution over possible communities.

\subsection{Probabilistic Generative Model}

Our data can be represented as an $N$ $\times$ $D$ matrix $\mbox{C}$ where each row represents a learner $n$ that has posted at least once in the course's online forums and each column $d$ represents a particular content label for each of the five dimensions described in the previous section (for example, one category in the Knowledge Construction dimension is ``statement of observation or opinion'').  Each entry of $\mbox{C}$, $c_{nd}$, is 1 if learner $n$ has made at least one post assigned a content label of $d$, and 0 otherwise.  Hence, $\mbox{C}$ is a binary matrix (our content analysis scheme allowed each post to be labelled with only one category per dimension -- however, users with multiple posts may be characterized by many different categories).  

$\mbox{C}$ depicts a bipartite \textit{learner-to-category} network.  Given our interest in uncovering latent groups of learners based on the category labels of their posts, we adopt the convention from (Psorakis et al. 2011b) and compute a standard weighted one-mode projection of $\mbox{C}$ onto the set of nodes representing learners, i.e. $\{n_{i}\}_{i=1}^{N}$.  The resultant $N$ $\times$ $N$ matrix $\mbox{X}$ has entries $x_{ij} = \sum_{d=1}^{D}{c_{id}c_{jd}}$, i.e., the total number of shared categories across all posts made by learners $i$ and $j$.  It is important to note that connections between learners $i$ and $j$ in this adjacency matrix do \textit{not} necessarily depict communication between them; instead, they indicate similar discussion contributions as defined by the content category labels assigned to their posts.  

We assume that the pairwise similarities described by $\mbox{X}$ are drawn from a Poisson distribution with rate $\hat{\mbox{X}}$ = $\mbox{WH}$, i.e. $x_{ij}$ $\sim$ $\mbox{Poisson}$($\sum_{k=1}^{K}{w_{ik}h_{kj}}$), where the inner rank $K$ denotes the unknown number of communities and each element $k$ for a particular row $i$ of $\mbox{W}$ and column $j$ of $\mbox{H}$ indicates the extent to which a single community contributes to $\hat{x}_{i,j}$.  In other words, the expected number of categories that two individuals $i$,$j$ share across their posts, $\hat{x}_{i,j}$, is a result of the degree to which they produce similar discussion content.  To address the fact that the number of communities $K$ is not initially known, we place \textit{automatic relevance determination} (MacKay, 1995) priors $\beta_{k}$ on the latent variables $w_{ik}$ and $h_{kj}$, similar to (Tan and F\`{e}votte, 2009), which helps ensure that irrelevant communities do not contribute to explaining the similarities encoded in $\mbox{X}$.  

The joint distribution over all the model variables is given by:

\begin{center}
\begin{tabular}{ll}
$p$(X,W,H,$\beta$) = $p$(X$\vert$W,H)$p$(W$\vert$$\beta$)$p$(H$\vert$$\beta$)$p$($\beta$) & (1)
\end{tabular}
\end{center}

And the posterior distribution over the model parameters given the data $\mbox{X}$ is:

\begin{center}
\begin{tabular}{ll}
$p$(W,H,$\beta$$\vert$X) = $\frac{p(\text{X} \vert \text{W,H})p(\text{W} \vert \beta)p(\text{H} \vert \beta)p(\beta)}{p(\text{X})}$ & \; \; \; (2)
\end{tabular}
\end{center}

\subsection{Inference and Cluster Assignment}

Our objective is to maximize the model posterior given the data $\mbox{X}$, which is equivalent to minimising the negative log posterior (i.e., the numerator of equation (2)) since $p$(X) is not a random quantity.  Like (Psorakis et al. 2011a), we represent the negative log posterior as an energy function {\cal U}:

\begin{center}
\begin{tabular}{llll}
{\cal U} & = & -- $\log{p(\mbox{X} \vert \mbox{W,H})}$ -- $\log{p(\mbox{W} \vert \beta )}$ \tabularnewline
& & -- $\log{p(\mbox{H} \vert \beta)}$ -- $\log{p(\beta)} \hspace{2.9cm} (3)$ 
\end{tabular}
\end{center}

The first term of {\cal U} is the log-likelihood of the data, $p$(X$\vert$W,H) = $p$(X$\vert$$\hat{X}$), which represents the probability of observing similar post content between two users $i$ and $j$ represented by $x_{ij}$, given an expected  (or Poisson rate) of $\hat{x}_{ij}$.  The negative log-likelihood is given by:

\begin{center}

\begin{tabular}{llll}
-- $\log{p(\mbox{X} \vert \hat{\mbox{X}})}$ & = & -- $\sum_{i=1}^{N}{\sum_{j=1}^{N}{\log{p(x_{ij} \vert \hat{x}_{ij})}}}$\tabularnewline\tabularnewline
                      & = & $\sum_{i=1}^{N}{\sum_{j=1}^{N}{\bigl(x_{ij}\log{\frac{x_{ij}}{\hat{x}_{ij}}}} + \hat{x}_{ij} }$ \tabularnewline
                      & & $ - x_{ij} + \frac{1}{2}\log{(2 \pi x_{ij})} \bigr) + \text{const.}$ (4)
\end{tabular}

\end{center}

Following (Tan and F\`{e}votte, 2009; Psorakis et al. 2011a), we place independent half-normal priors over the columns of $W$ and rows of $H$ with zero mean and precision (inverse variance) parameters $\beta$ $\in$ $\mathbb{R}^{K}$ = [$\beta_{1}$,...,$\beta_{K}$].  The negative log priors are:  

\begin{center}
\begin{tabular}{llll}

-- $\log{p(\mbox{W} \vert \beta)}$ & = & -- $\sum_{i=1}^{N}{\sum_{k=1}^{K}{\log{{\cal HN}(w_{ik}; 0,\beta_{k}^{-1})}}}$\tabularnewline\tabularnewline
						 & = & $\sum_{i=1}^{N}{\sum_{k=1}^{K}{(\frac{1}{2}\beta_{k}w_{ik}^{2}) - \frac{N}{2}\log{\beta_{k}}}}$\tabularnewline
						 & & + $\text{const.}$ \hspace{3.1cm} (5) \tabularnewline\tabularnewline
-- $\log{p(\mbox{H} \vert \beta)}$ & = & -- $\sum_{k=1}^{K}{\sum_{i=1}^{N}{\log{{\cal HN}(h_{ki}; 0,\beta_{k}^{-1})}}}$\tabularnewline\tabularnewline
						 & = & $\sum_{k=1}^{K}{\sum_{i=1}^{N}{(\frac{1}{2}\beta_{k}h_{ki}^{2}) - \frac{N}{2}\log{\beta_{k}}}}$\tabularnewline
						 & & + $\text{const.}$ \hspace{3.1cm} (6)\tabularnewline\tabularnewline
\end{tabular}
\end{center}

Each $\beta_{k}$ controls the importance of community $k$ in explaining the observed interactions; large values of $\beta_{k}$ denote that the elements of column $k$ of $\mbox{W}$ and row $k$ of $\mbox{H}$ lie close to zero and therefore represent irrelevant communities.  Further to (Psorakis et al. 2011a), we assume the $\beta_{k}$ are independent and place a standard Gamma distribution over them, yielding the following negative log hyper-priors:

\begin{center}
\begin{tabular}{llll}

- $\log{p(\beta)}$ & = & -- $\sum_{k=1}^{K}{\log{{\cal G}(\beta_{k}\vert a,b)}}$\tabularnewline\tabularnewline
						 & = & $\sum_{k=1}^{K}{(\beta_{k}b - (a - 1) \log{\beta_{k})}}$\tabularnewline
& & + \text{const.} \hspace{3.5cm} (7) 
\end{tabular}
\end{center}

The objective function {\cal U} of Eq. (3) can be expressed as the sum of Equations (4), (5), (6), and (7).  To optimize for $\mbox{W}$, $\mbox{X}$, and $\beta$, we use the fast fixed-point algorithm presented in (Tan and F\`{e}votte, 2009; Psorakis et al., 2011a) with algorithmic complexity $\mbox{O}(NK)$, which involves consecutive updates of $\mbox{W}$, $\mbox{H}$, $\beta$ until convergence (e.g. a maximum number of iterations) has been satisfied.  The pseudocode for this procedure is presented in Figure \ref{comm-det-bnmf}.  

\begin{figure}[t]
\centering
\begin{algorithm}{CD-BNMF}{\mbox{X}; \ K_{0}; \ a; \ b}
\textbf{for} \; i=1 \; to \; n_{iter} \; \textbf{do}\\
\hspace{.5cm}\mbox{H} \gets (\frac{\mbox{H}}{\mbox{W}^{\top}1+\beta \mbox{H}})\mbox{W}^{\top}(\frac{\mbox{X}}{\mbox{WH}})\\
\hspace{.5cm}\mbox{W} \gets (\frac{\mbox{W}}{\mbox{1H}^{\top}1+\mbox{W} \beta})(\frac{X}{\mbox{WH}}) \mbox{H}^{\top}\\
\hspace{.5cm}\beta_{k} \gets \frac{N + a - 1}{\frac{1}{2}(\sum_{i}{w_{ik}^{2}} + \sum_{j}{h_{kj}^2})+b}\\
\textbf{end for}\\
K_{*} \gets \; \# \; nonzero \; columns \; of \; \mbox{W} \; or \; rows \; of \; \mbox{H}\\
\textbf{return} \; \mbox{W}_{*} \in \mathbb{R}_{+}^{N \times K_{*}}, \mbox{H}_{*} \in \mathbb{R}_{+}^{K_{*} \times N}
\end{algorithm}
\caption{Community Detection using Bayesian NMF.}
\label{comm-det-bnmf}
\end{figure}

In the case of our application, $\mbox{W}_{*}$ = $\mbox{H}_{*}^{\top}$ since $\mbox{X}$ is symmetric.  Each element $w_{ik}^{*}$, or $h_{ki}^{*}$ denotes the \textit{degree of participation} of individual $i$ in cluster $k$ while each normalized row of $\mbox{W}_{*}$ (or column of $\mbox{H}_{*}$) expresses each node's \textit{soft-membership} distribution over the possible clusters.  This soft-membership provides more context to our \textit{belief} about a node's cluster membership, which we can model and explore explicitly if desired. 

\section{Results}

\subsection{Model Benchmarking}

In order to evaluate the BNMF model and inference scheme's ability to represent the data, we computed the negative log likelihood (NLL) and root mean-squared error (RMSE) of held-out test data against the nonparametric linear Poisson gamma model (LPGM) of (Gupta et al., 2012).  The LPGM is a latent feature model that treats the number of hidden clusters as a variable to be learned during inference, leveraging an Indian Buffet Process prior over a infinite-dimensional binary hidden feature matrix.  The LPGM model specifies: 

\begin{center}
\begin{tabular}{lll}

$\mbox{X}$ & = & $(\mbox{Z} \circ \mbox{F}) \mbox{T} + \mbox{E}$

\end{tabular}
\end{center}

Here, $\mbox{X}$ is the $N$ $\times$ $D$ matrix of observations; $\mbox{Z}$ is the $N$ $\times$ $K$ binary latent feature matrix with entry $i,k$ = 1 iff feature $k$ is represented in datum (i.e., learner) $i$; $\mbox{T}$ is a non-negative $K$ $\times$ $D$ matrix illustrating the represenation of each dimension $d \in D$ in feature $k \in K$; $\mbox{F}$ is an $N$ $\times$ $K$ non-negative matrix indicating the strength of participation of $i$ in feature $k$; $\mbox{E}$ is the reconstruction error with rate $\lambda$ such that $\mbox{E}_{ij}$ $\sim$ $\mbox{Poisson}$($\lambda$); and $\circ$ is the Hadamard element-wise product operator.

Given the scalability issues of the IBP, we evaluated both BNMF and LPGM on 20 randomly-selected 50 $\times$ 50 subsets of real-world data by comparing the root mean squared error (RMSE) and negative log likelihood (NLL) of held-out test data.  Each 50 $\times$ 50 subset was determined by randomly sampling the rows and corresponding columns of real-world content-analysed data, and for each row, 10\% of entries were randomly selected for hold-out, with the remainder used for training.  Inference on the infinite model was performed via 5000 iterations of Gibbs sampling, with no samples discarded for burn-in.  For comparison purposes, we computed the RMSE and NLL of a na\"{i}ve model (Pred-Avg) that predicts the arithmetic mean of the training data for each held-out data point, as well as the RMSE for a na\"{i}ve benchmark (Pred-0) that always predicts 0 for all held-out data (because of the 0 prediction, computing the NLL for this na\"{i}ve model would involve repeatedly evaluating a Poisson likelihood with rate $\lambda$ = 0, ultimately yielding $\infty$).  Table \ref{bnmf-ibp-compare}  summarizes the results, which reveal that the proposed BNMF model and inference scheme has greater predictive accuracy than its nonparametric IBP counterpart (confined to a finite number of sampling iterations) and both na\"{i}ve approaches\footnote{the values presented in the table are generated by taking the arithmetic mean of the RMSE and NLL, computed for each of the 20 different subsets, with different data held-out for each subset.}.  Moreover, given its computational tractability (taking seconds to run on the full dataset, versus days for the LPGM), BNMF offers a favourable generative model for extracting latent features from online discussion forum data. 

\begin{table}[t]
\begin{center}
\begin{tabular}{@{}lllll@{}}
\toprule
 & BNMF & LPGM & Pred-Avg & Pred-0 \tabularnewline
\midrule
RMSE & \textbf{0.4647} & 1.2199 & 0.9330 & 1.5823\tabularnewline 
\hline
NLL & \textbf{251.97} & 355.22 & 324.74 & --\tabularnewline
\bottomrule
\end{tabular}
\caption{RMSE and NLL results for the BNMF, LPGM, Pred-Avg, and Pred-0 models.  Bold values indicate the strongest predictive performance on held-out test data.}
\label{bnmf-ibp-compare}
\end{center}
\end{table}

\subsection{Exploring Extracted Communities}
\vspace{-1mm}
Students in the business strategy course were encouraged to interact through its online discussion forum, which was segmented into multiple sub-forums.  Two sub-forums aimed at promoting learner engagement and interactions were content-analysed and explored for latent features:  Cases and Final Projects.  The Cases sub-forum facilitated weekly discussions about a real company and its business challenges/opportunities (for example, in week 1, the selected company was Google, and one of the questions was ``Do you think Google's industry is a competitive market, in the technical sense? Does Google have a sustainable competitive advantage in internet search?'').  The Final Project sub-forum facilitated questions, debates, and team formation for the final strategic analysis assignment.  The remaining sub-forums were:  Questions for Professor, Technical Feedback, Course Material Feedback, Readings, Lectures, and Study Groups.

Since participation in multiple sub-forums was minimal (in most cases, no more than 10\% of participants in one sub-forum participated in another), we explored the latent features of communication and the characteristics of these underlying communities in both sub-forums independently of one another.

Learners were assigned to inferred communities by computing maximum a-posteriori (MAP) estimates for $\mbox{W}$ and $\mbox{H}$ as described in section 2 and greedily assigning each learner $i$ to the community $k_{i}^{*}$ to which it ``most'' belongs, i.e. $k_{i}^{*}$ = $\mbox{argmax}_{k \in K} w_{ik}$.  In repeated executions of the BNMF procedure, different community assignments were computed for some learners due to random initializations of $\mbox{W}$ and $\mbox{H}$ as well as numerical precision issues that affected the group allocation step.  To mitigate this, we ran the algorithm 100 times and used the factor matrices with the highest data likelihood to compute the final allocations.

Our analysis of extracted communities sought to understand the demographics, course outcomes, broader forum behaviours and types of posts for each of its constituent learners. 

\subsubsection{Cases sub-forum}
The Cases sub-forum had 1387 unique participants that created nearly 4,100 posts or comments.  We used BNMF to detect latent communities based on the learning and dialogue acts reflected in these posts, as this particular sub-forum was set up for participants to practice the tools and frameworks they learned in the course, and so, the learning and dialogue dimensions were selected to reveal the ways in which people used the forums to engage with one another and construct knowledge.  Four learner communities emerged, containing 238, 118, 500, and 531 people, respectively.  We describe these communities as committed crowd engagers, discussion initiators, strategists, and individuals, respectively.  

\textbf{Community 1 (committed crowd engagers)}. Participants in this group tended to engage with others in the forum.  Of all the groups they contributed the most responsive dialogue acts at 43\% of total posts, and the second highest number of informative (8\%) and elecitive (5\%) statements. In terms of learning, they tended to achieve quite similar levels of higher-order knowledge construction to groups 2 and 3. These participants read and posted the most of all four groups. 45\% of the group's participants passed the course -- significantly more than any other group (p$\textless$0.05
\footnote{We used the nonparametric Kruskal-Wallis one-way analysis of variance to test for statistical significance.}). Interestingly, members of this group were likely to be from Western continents, with a larger proportion of Europeans (26.1\%), albeit only significantly greater than the other groups at the p$\textless$0.1 level.  Nearly 31\% had at least a Master's degree -- similar to group 3. It is reasonable to suggest that this group found the forums an important part of their learning and used it as they sought to formally pass the course.

\textbf{Community 2 (discussion initiators)}.  Most notable for this group was its level of elicitative dialogue acts -- which characterized over 48\% of its participants' posts. Morever, 24\% of their posts did not involve learning, a significantly greater proportion than the other groups (p$\textless$0.07 compared to group 1; p$\approx$0 compared to groups 3 and 4).  Still, members of this group had a larger proportion of posts reflecting higher-order learning than the other groups (8.0\%).  Interestingly, this group had a significantly lower pass rate than groups 1 and 3 (25\%, p$\textless$0.05), but this could be explained to a large extent by the high number of people who did not submit a final project (67\%, similar to group 4).  Members of this group viewed fewer discussion threads and contributed fewer posts than groups 1 and 3. Geographically speaking, a significantly higher proportion of this group's members were located in Asia in comparison to the other three (31\%, p$\textless$0.01).  This could suggest that geography played an important role in motivating discussion.  Indeed, the more elicitative nature of dialogue in this group may suggest cultural differences in interpretations of, or responses to, various conversation topics. 

\textbf{Community 3 (strategists)}.  In many ways, people in this group were similar to group 1. They had similar levels of higher-order learning and tended to be responsive to others' comments. However, they had a greater proportion of argumentative statements (55\%) and rarely had posts that reflected no learning (1.6\%). People in this group were second most likely to pass the exam (36.2\%) and second most likely to try to pass, but ultimately fail (6.4\%). They tended to be similarly educated to those in group 1 -- with nearly 30\% receiving at least a Master's degree. They viewed and contributed to the forums the second most number of times, but this was still significantly less than group 1 (p $\approx$ 0).  Combined, these characteristics suggest that students in group 3 were more strategic in their approaches, using the Cases sub-forum only as needed to achieve their learning goals.  

\textbf{Community 4 (individualists)}.  People in group 4 were highly distinctive in their large proportion of argumentative statements (85\%). They had a smaller proportion of posts featuring higher-order learning (3.7\%) compared to groups 1 - 3.  They read and posted in the forums less than any other group (significant at p$\approx$0 compared to groups 1 and 3).  They were the most likely to not submit a final project (68\%) – a similar number to group 2. Of all the groups, participants in this group had the smallest proportion of people attain at least a Master's degree (23.2\%, p$\textless$0.05 compared to groups 1 and 3). These indicators may suggest a number of possibilities:  that members of this group were the most likely to drop out of the course of all four groups, may have had limited experience of using forums to construct their knowledge, or simply preferred to learn individually.

Figure 2 shows the dialogue acts and geographic locations of the members of each group.

\begin{figure*}[htp]
  \hspace{-1cm}
  \subfigure[]{\includegraphics[width=9cm]{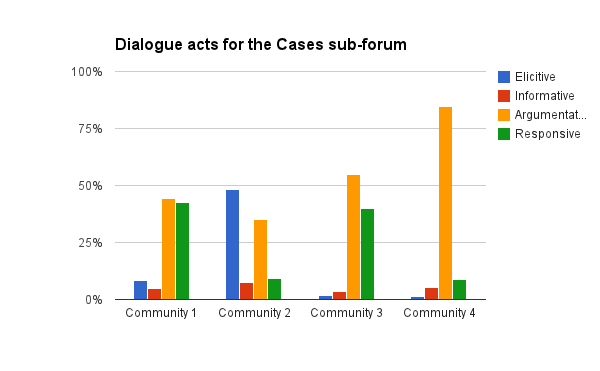}}\quad
  \subfigure[]{\includegraphics[width=9cm]{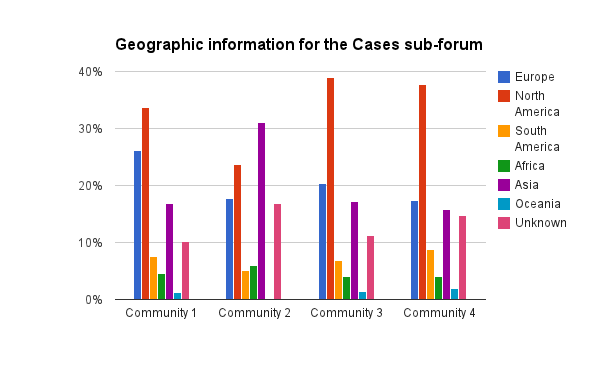}}
  \caption{Plot (a) illustrates the dialogue acts represented in the posts made by learners belonging to each community, and plot (b) depicts their geographic locations.  The geographies of some learners are ``unknown'' due to technical errors in data collection in the online learning environment.}
\end{figure*}

\subsubsection{Final Project sub-forum}
The Final Projects sub-forum had 1256 unique participants creating nearly 2,400 posts or comments.  We selected the communication and topic labels as inputs into BNMF because of the nature of the sub-forum:  it was a place for participants to find others to discuss their individual final projects with prior to the submission.  Therefore, how people engaged with each other and the topics of their engagements were central to this setting.  We detected 5 communities with 296, 50, 611, 45, and 237 individuals, which we characterised as:  instrumental help seekers, careful assessors, community builders, focused achievers, and project support seekers, respectively.  \footnote{17 individuals were assigned to their own groups; for the purposes of analysis, we only investigated clusters with at least two members.}

\textbf{Community 1 (instrumental help seekers)}.  Participants in this group had a high proportion of elicitative dialogue acts (64\%) and primarily discussed the final project (83\%).  On average, they posted more than groups 2 and 4 and their amount of views of the forum were relatively low (similar to groups 4 and 5).  The proportion of people who passed was significantly lower than in groups 2, 3 and 4 (41\%, p$\textless$0.01).  People in this group were also more likely to submit and fail the final project than clusters 3 and 5 (14\%, p$\leq$0.05).  There were fewer people with postgraduate qualifications compared to groups 3 and 5 (20\%, p$\textless$0.01). These trends suggest that members of this community sought help by asking questions and discussing the final project with their peers, but still did not pass the course.

\textbf{Community 2 (careful assessors)}. Participants in this group had the highest proportion of elicitative dialogue acts out of all of the groups (71\%), but in contrast to group 1, the focus of their posts was about the peer review process (87\%). They viewed more posts on average than groups 1, 4, and 5, but only groups 4 posted fewer comments on average.  Thus, it seems that participants in this group used the forums to look for answers to questions they had about peer review, and only posted again if necessary.   Like group 4, a high proportion of learners passed the course, compared to groups 1, 3, and 5 (p$\textless$0.05).  These patterns suggest that this group needed to know more about the peer assessment process, but that its members were very strategic in their use of this sub-forum to obtain necessary information. 

\textbf{Community 3 (community builders)}. Participants in this group were distinctive in the proportion of posts that were responsive to others (55\%). In contrast to the other groups, the focus of their discussions were spread across final projects and peer review.  Interestingly this group seemed the most engaged of groups in the forum, being the most likely to view and post in this sub-forum of all participants in other groups (p$\textless$0.05, p$\textless$0.001, respectively). Likewise, the average length of posts submitted by supporters (712 words) was markedly higher than in any other group (Group 1 had the 2nd highest average of 382 words -- p$\textless$0.001). Their pass rate (51\%) was higher than clusters 1 and 5 (p$\textless$0.01), but lower than 2 and 4 (p$\textless$0.05), partly due to the high proportion of learners (41\%) that did not submit a final project. This suggests that participants in group 3 were more interested in exchanging ideas with others as opposed to receiving formal acknowledgement or recognition for passing the course. 

\textbf{Community 4 (focused achievers)}. Participants in this group were distinctive as they had a higher proportion of argumentative dialogue acts (68\%). While most focus was on peer review (70\%), many posts also discussed course outcomes and certificates (20\%). They had the highest proportion of posts that evidenced some form of learning (32\%).  They posted the least (on average, 2.5 times), and had the smallest average post size (146 words) and number of thread views views (38), both statistically significant only when compared to group 3 (p$\textless$0.05). Interestingly, they had the highest proportion of participants submit a final project and pass the course (76\%, p$\textless$0.01 compared to groups 1, 3, and 5), yet a similar proportion to group 1 who submitted but still failed (13\%). Furthermore, they comprised a group that showed the most emotion in their posts (20\%) of all the groups. These patterns suggest a very focused group of participants who only used the forums when necessary to achieve their goals -- and to express both joy and unhappiness with their own course outcomes.

\textbf{Community 5 (project support seekers)}. Participants in this group were similar to those in group 1, although they were distinguished by a high proportion of imperative dialogue acts (50\%) and organizing virtual meet-ups (45\%). The average number of discussion thread views was relatively low (40 - similar to groups 1 and 4); moreover, participants made posts more often than groups 2 and 4, albeit not with statistical significance. This pattern suggests that participants in this group were seeking support and opportunities for collaboration on the final project.  Interestingly, this was the only group where significant differences were found in geographic region:  there were more people from South America in this group compared to 3 (p$\textless$0.01), which may indicate a wish for people from the same part of the world to collaborate. While this group had a higher number of participants with postgraduate degrees than group 1 (29\%, p$\textless$0.05), they had the lowest pass rate out of all other groups (32\%, p$\textless$0.05), partly explained by having the highest proportion of participants who did not submit a final project (57\%, p$\textless$0.05).

Figure 3 shows the discussion topics and course outcomes of the members of each group.

\begin{figure*}[htp]
  \hspace{-1cm}
  \subfigure[]{\includegraphics[width=9cm]{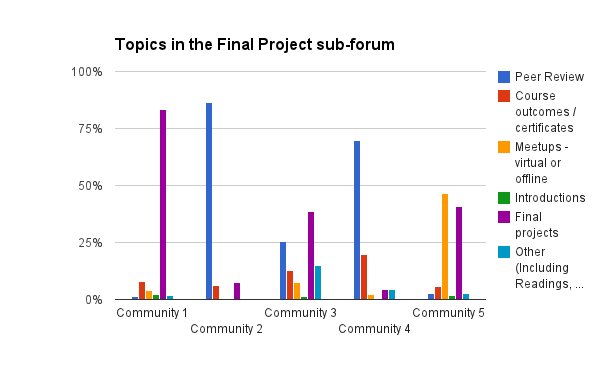}}\quad
  \subfigure[]{\includegraphics[width=9cm]{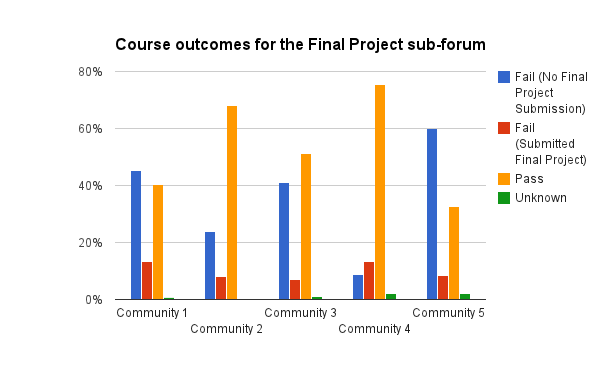}}
  \caption{Plot (a) illustrates the discussion topoics represented in the posts made by learners belonging to each community, and plot (b) depicts their course outcomes.  The outcomes of some learners are ``unknown'' due to technical errors in data collection in the online learning environment.}
\end{figure*} 

\section{Discussion}

Using BNMF to extract latent features from the dataset and subsequently exploring the composition of these features reveals that the different sub-forums in MOOCs offer participants different ways to engage with course content -- and each other.  In our analysis, we see that the cases sub-forum is actively facilitating subject specific learning to some degree, and the final projects sub-forum is more geared towards social, instrumental and practical aspects of the learning process.  The distinctively different interaction patterns that characterize the groups in each sub-forum indicate that learners have very different needs and expectations of the discussion forums, and these needs must be considered in order to truly understand how to support learning in massive open online courses.

The different profiles or communities identified in each of the forums relate very well to existing educational research.  Discussion forums encourage participants to enhance their understanding through discussing specific questions with others.  Participants can, to a large extent, choose for themselves how much they wish to use the forums to construct knowledge together, i.e. adopting a more socio-cultural approach to learning, or use the forums as a way to reflect on their own ideas, more in-line with cognitive and social constructivist approach to learning (Stahl, 2006). In the four communities observed within the Cases sub-forum, we see this from the most individual learners (individualists) to the most collaborative (committed crowd engagers).  Similarly, in the Final Projects sub-forum, we see people leveraging the crowd in different ways to support their information-seeking and help-seeking behaviours. Groups 1 (instrumental help seekers) and 5 (project support seekers) in particular turn to the crowd for support. Learners in Group 3 (community builders) use the forum to enhance their own expertise and identity by exchanging ideas with others. Those in group 2 (careful assessors) view the forum as a more informational -- effectively, a text-based -- resource. The sheer scale of the forums and confusion about expectations may have encouraged learners to engage in a multitude of ways -- producing an eclectic mix of learning theories in action. 

It is not possible to use the data and corresponding results to make judgements about which of our profiles lead to the most positive learning outcomes, as traditional educational outcomes may not necessarily be used to understand MOOCs (Kizilcec et al., 2013). Yet, it is clear that different groups may need varying kinds of support in order to help them achieve their learning goals (however they define them). 

For example, the individualists in the Cases forum may simply prefer to learn individually and just use the group as a way to present their ideas to the world, or they may have had challenges in engaging fully within the forums.  If it is the latter, then this group requires additional support in order to learn through discussions.  The fact that this group has a lower proportion of postgraduate qualifications relative to other groups and lower pass rates may warrant further exploration.  Similarly, in the Final Projects forum the instrumental help seekers asked many questions about the final projects, yet given the relatively low proportion of people who passed, it appears that these requests for help were unrequited. Thus, clearer information -- or ways of accessing information about assessments -- would be beneficial for this group. 

There are a number of cultural trends that emerge from the data that warrant further consideration.  For example, in the Cases sub-forum, group 2 (discussion initiators) contained a relatively high proportion of people from Asia. This suggests that people in the same regions tend to communicate with one another in a similar fashion. In the same vein, a relatively high proportion of group 5 (project support seekers) in the Final Projects sub-forum -- who used the discussion space to seek groups for the final assignment -- were from South America.  While these continental regions are massive in their own right, geographic discrepancies in the inferred communities further suggest the potential influence of cultural differences, perspectives, and preferences in educational contexts.  Exploring these influences may enable researchers and practitioners to better facilitate local connections and meet-ups in the otherwise global-scale online learning setting.

Finally, it is important to note the flexibility and potential for practical implementation afforded by our computational toolset of choice.  BNMF's specification of a robust probabilistic generative model capable of representing the content-labelled data well, as validated against the LPGM and other benchmarks, makes it a prime candidate to serve as a community detection (and perhaps, simulation) tool geared towards detecting posting behaviours and latent learner groups in online courses.  Coupled with its low computational overhead, it is possible that in future learning platforms, algorithms like BNMF could run on cloud-based services in real-time during a course to deliver valuable insights to course staff and students, ultimately improving learning experiences and outcomes. 

\section{Conclusion}

This paper demonstrates how latent feature models can reveal underlying structures in the content of discussions in massive open online courses.  We introduced a new content-analysed set of MOOC forum data and used Bayesian Non-negative Matrix Factorization with a computationally efficient and highly scalable inference scheme to cluster users based on the learning, dialogue acts, affect, topic, and local relevance of their posts.  By exploring the underlying user groups in two sub-forums with relatively high levels of engagement and interaction, we uncovered statistically significant differences in the demographics, post content, course outcomes, and engagement of learners within each group -- enabling the creation of learner profiles that characterise discussion forum use.  These findings are not only important because of their implications for the Education community, but also because they demonstrate how principled probabilistic generative modelling and inference can enable the Machine Learning community to explore, and model, nuanced, large-scale learner data from MOOCs.  The ability to develop strong generative models and computationally efficient inference schemes for educational data promises to advance insights in online learning that can help improve course design and pedagogical practices.  These scalable and robust models will be crucial inputs into the learning platforms of the future, equipped to cater to the needs of a diverse, global body of lifelong learners.  

\subsubsection*{Acknowledgements}

The authors would like to thank the MOOC Research Initiative for enabling this interdisciplinary work, as well as the University of Virginia and Coursera for their support in accessing relevant datasets.  Thanks also to Chris Davies and Bav Radia in Oxford's Department of Education for their help in content analysis, and Taha Yasseri, Ioannis Psorakis, Rory Beard, Tom Gunther, and Chris Lloyd for their insights.  Finally, thanks to Sunil Gupta at the University of Deakin for sharing code for the LPGM.


\subsubsection*{References}

A. Anderson, D. Huttenlocher, J. Kleinberg, J. Leskovec (2014).  Engaging with Massive Open Online Courses.  {\it Proceedings of the 14th International World Wide Web Conference Committee}.  Seoul, Korea.  

B. De Weaver, T. Schellens, M. Valcke, and H. Van Keer (2006) Content analysis schemes to analyze transcripts of online asynchronous discussion groups: A review.  {\it Computers and Education}, 46(1): 6-28.

C. G. Brinton, M. Chiang, S. Jain, H. Lam, Z. Liu, F. M. F. Wong (2013).  Learning about social learning in MOOCs:  From statistical analysis to generative model.  arXiv:1312.2159v2.

C. N. Gunawardena, C. A. Lowe, and T. Anderson (1997). Analysis of a Global Online Debate and the Development of an Interaction Analysis Model for Examining Social Construction of Knowledge in Computer Conferencing. {\it Journal of Educational Computing Research}, 17(4):  397–431.

C. Reed and Z. Ghahramani (2013).  Scaling the Indian Buffet Process via Submodular Maximization. arXiv:1304.3285v4.

C. Rose (2013).  Enabling Resilient Massive Scale Open Online Learning Communities through Models of Social Emergence.  {\it MOOC Research Initiative Conference}, Arlington, TX. 

D. B. Clark, V. Sampson, A. Weinberger, and G. Erkens (2007). Analytic Frameworks for Assessing Dialogic Argumentation in Online Learning Environments. {\it Educational Psychology Review}, 19(3): 343–374. 

D. Blei, A. Ng, M. Jordan (2003).  Latent Dirichlet Allocation.  {\it Journal of Machine Learning Research}, 3: 993-1022. 

D. D. Lee, H. S. Seung (1999).  Learning the parts of objects by non-negative matrix factorization.  {\it Nature}, 401: 788-791.  doi:10.1038/44565.

D. F. Nobel (1998).  Digital diploma mills:  the automation of higher education.  {\it Science as Culture}, 7(3): 355-368.

D. J. C. Mackay (1995). Probable networks and plausible predictions a review of practical Bayesian models for
supervised neural networks. {\it Network: Computation in Neural Systems}, 6(3):469–505.

E. Airoldi, D. M. Blei, S. E. Feinberg, E. P. Xing (2008).  Mixed Membership Stochastic Blockmodels.  {\it Journal of Machine Learning Research}, 9: 1981-2014.  

F. Doshi-Velez (2009).  {\it The Indian Buffet Process:  Scalable Inference and Extensions} (Master's Dissertation).  University of Cambridge, Cambridge, UK.    

G. Erkens, and J. Janssen (2008). Automatic coding of dialogue acts in collaboration protocols. {\it International Journal of Computer-Supported Collaborative Learning}, 3(4): 447–470. doi:10.1007/s11412-008-9052-6

G. Stahl, T. Koschmann and D. Suthers (2006). Computer-supported collaborative learning: An historical perspective. In R. K. Sawyer, ed. {\it Cambridge handbook of the learning sciences}.  Cambridge, UK: Cambridge University Press, 409-426. 

H. Xiao, T. Stibor (2010).  Efficient Collapsed Gibbs Sampling For Latent Dirichlet Allocation.  {\it Journal of Machine Learning Research}, 13: 63-78.

I. Holmes, K. Harris, C. Quince (2012) Dirichlet Multinomial Mixtures: Generative Models for Microbial Metagenomics. PLoS ONE 7(2): e30126. 

I. Psorakis, S. Roberts, M. Ebden, B. Sheldon (2011a).  Overlapping Community Detection using Nonnegative Matrix Factorization.  {\it Physical Review E}, 83, 066114.

I. Psorakis, S. Roberts, I. Rezek and B. Sheldon (2011b).  Inferring social network structure in ecological systems from spatio-temporal data streams. {\it Journal of the Royal Society Interface}, 9(76):  3055-3066.

L. Vaquero and M. Cebrian (2013).  The rich club phenomenon in the classroom.  {\it Nature: Scientific Reports}, pp. 1-8.

L. Vygotsky (1978).  {\it Mind in Society}.  Harvard University Press, Cambridge, MA.

M. E. J. Newman, M. Girvan (2003).  Finding and evaluating community structure in networks.  arXiv:cond-mat/0308217v1.

M. K. Titsias (2007).  The Infinite Gamma-Poisson Feature Model.  {\it Advances in Neural Information Processing Systems 20}, 1513-1520.  Vancouver, B.C.

M. N. Schmidt, O. Winther, L. K. Hansen (2009).  Bayesian Non-negative Matrix Factorization.  {\it Independent Component Analysis and Signal Separation Lecture Notes in Computer Science}, 5441: 540-547.

R. Kizilcec, C. Piece, Schneider, E. (2013).  Deconstructing Disengagement:  Analyzing Learner Subpopulations in Massive Open Online Courses.  {\it The 3rd Proceedings of the Learning Analytics and Knowledge Conference}, Leuven, Belgium.

R. Pekrun, T. Goetz, W. Titz, and R. P. Perry (2002). Academic Emotions in Students’ Self-Regulated Learning and Achievement: A Program of Qualitative and Quantitative Research. {\it Educational Psychologist}, 37(2), 91–105.

S. Geman and D. Geman (1984).  Stochastic Relaxation, Gibbs Distributions, and the Bayesian Restoration of Images.  {\it IEEE Transactions on Pattern Analysis and Machine Intelligence}, 6(6): 721-741. 

S. K. Gupta, D. Phung, S. Venkatesh (2012).  A Nonparametric Bayesian Poisson Gamma Model for Count Data.  {\it Proceedings of 21st International Conference on Pattern Recognition}, 1815-1818.  Tsubka Science City, Japan.

T. Griffiths and Z. Ghahramani (2005).  Infinite latent feature models and the Indian buffet process.  {\it Technical Report 2005-001, Gatsby Computational Neuroscience Unit}. 

V. Tan and C. F\`{e}votte (2009). Automatic relevance determination in nonnegative matrix factorization. {\it SPARS09 - Signal Processing with Adaptive Sparse Structured Representations}, 1–19.

\end{document}